# Induced Superconductivity in Hybrid Au/YBa$_2$Cu$_3$O$_{7-x}$ Electrodes on Vicinal Substrates

Irina I. Gundareva[1,3*], Jose Martinez-Castro[2,3,4], F. Stefan Tautz[2,3,5], Detlev Grützmacher[1,3], Thomas Schäpers[1,3], Matvey Lyatti[1,3]

[1] Peter Grünberg Institute (PGI-9), Forschungszentrum Jülich, 52425 Jülich, Germany

[2] Peter Grünberg Institute (PGI-3), Forschungszentrum Jülich, 52425 Jülich, Germany

[3] Jülich Aachen Research Alliance, Fundamentals of Future Information Technology, 52425 Jülich, Germany

[4] Institute of Experimental Physics II B, RWTH Aachen, 52074 Aachen, Germany

[5] Institute of Experimental Physics IV A, RWTH Aachen, 52074 Aachen, Germany.

[*] i.gundareva@fz-juelich.de

**Abstract**

Superconducting electrodes are an integral part of hybrid Josephson junctions used in many applications including quantum technologies. We report on the fabrication and characterization of superconducting hybrid Au/YBa$_2$Cu$_3$O$_{7-x}$ (YBCO) electrodes on vicinal substrates. In these structures, superconducting CuO$_2$-planes face the gold film, resulting in a higher value and smaller variation of the induced energy gap compared to the conventional Au/YBCO electrodes based on films with the **c**-axis normal to the substrate surface. Using scanning tunneling microscopy, we observe an energy gap of about 10-17 meV at the surface of the 15- nm-thick gold layer deposited *in situ* atop the YBCO film. To study the origin of this gap, we fabricate nanoconstrictions from the Au/YBCO heterostructure and measure their electrical transport characteristics. The conductance of the nanoconstrictions shows a series of dips due to multiple Andreev reflections in YBCO and gold providing clear evidence of the superconducting nature of the gap in gold. We consider the Au/YBCO electrodes to be a versatile platform for hybrid Josephson devices with a high operating temperature.

In recent years, several emerging 1D (one-dimensional) and 2D (two-dimensional) materials such as semiconductor nanowires, graphene, or topological insulators have been attracting growing attention. Although these materials possess unique properties, most of them obey the Fermi-Dirac statistics which does not allow for a macroscopic quantum state to appear. However, when they are combined with a superconductor, the macroscopic quantum state can be induced by the proximity effect close to the interface.[1] Having great potential for quantum technologies, such hybrid structures are also exciting objects for fundamental research.

For example, graphene proximitized with a superconductor is a promising platform for studying 2D quantum phase transitions, topological superconductivity, and hybrid devices for quantum computing.[2, 3] Furthermore, the interface between a superconductor and a topological insulator is predicted to host Majorana fermions, and many efforts have been made in search of Majorana particle that paves the way to the realization of a topologically protected fault-tolerant quantum computer.[4] The topological state can be realized using an s-wave superconductor coupled to a 1D semiconductor nanowire with a strong spin-orbit coupling and a high g-factor.[5-7] Semiconductor nanowires can also be used as weak links to create Josephson junctions for qubits based on other physical principles such as gatemons or Andreev-level qubits.[8] However, the value of the induced superconducting gap in hybrid devices based on low-$T_c$ superconductors is rather small and typically has μeV scale which makes them vulnerable to external interferences.

It has been theoretically predicted that an alternative approach using high-temperature (high-$T_c$) superconductors with $d_{x2-y2}$-wave symmetry of the order parameter and large anisotropic energy gaps of tens of meV could give far more powerful results.[9-13] The use of high-$T_c$ superconductors can not only increase the operating temperature and stability of hybrid devices but reveals exciting physics. First reports on graphene proximitized with a $Pr_{2-x}Ce_xCuO_4$ superconductor have shown signatures of proximity-induced unconventional p-wave pairing in graphene[14] and are in good agreement with theoretical predictions.[15] Several works are devoted to hybrid structures based on $Bi_2Sr_2CaCu_2O_8$ (BSCCO) covered by a thin layer of topological insulators $Bi_2Se_3$ or $Bi_2Te_3$.[16-21] Notably, the investigations of the hybrid structures by angle-resolved photoemission spectroscopy (ARPES) and scanning tunneling microscopy (STM) have given controversial results and the question of whether one can observe an induced gap in the topological insulators on top of a BSCCO superconductor is still under discussion.

The attempts to proximitize graphene with $YBa_2Cu_3O_{7-x}$ (YBCO) superconductor deposited on the conventional substrates showed only hints of induced superconductivity.[22, 23] On the one hand, as shown for example for a graphene/YBCO interface,[22, 23] many emerging materials for the hybrid Josephson junctions have poor compatibility with oxide superconductors. Direct coupling of these proximitized materials to the high-$T_c$ superconductor leads to an inferior interface with low transparency and correspondingly, a large drop of the induced energy gap $\Delta_{ind}$ in the interface layer (upper panel in Figure 1a). A mediating layer between the proximitized material and the oxide superconductor can solve this problem. The mediating layer has to provide a small order parameter attenuation and transparent interfaces to both of the contacting materials resulting in the larger induced energy gap in the proximitized material (lower panel in Figure 1a). On the other hand, similar to the other cuprate superconductors, the YBCO has a layered structure with superconducting $CuO_2$ planes, as shown in Figure 1b. The larger the contact area between the $CuO_2$ planes and the proximitized layer, the higher the magnitude of the induced superconducting gap is expected. All earlier works on hybrid devices with high-$T_c$ superconducting electrodes are based on the films where the **c**-axis is normal and the $CuO_2$ planes are parallel to the film surface (so-called **c**-axis oriented films).[24, 25] The sketch of such an electrode is shown in Figure 1c. The mediating layer in such electrodes is coupled to the CuO-chains where the order parameter is significantly reduced compared to $CuO_2$-planes. Therefore, the induced energy gap in these devices is expected to be rather small compared to the intrinsic YBCO energy gaps. Finite surface roughness of the films may provide a small contact area between the $CuO_2$ planes and the mediating layer that results in the small and strongly spatially inhomogeneous induced energy gap as experimentally shown for Au/YBCO heterostructures.[26] Such a configuration of the electrodes gives only a hint of induced superconductivity in proximitized material but could hardly ever allow the development of a reproducible hybrid device.

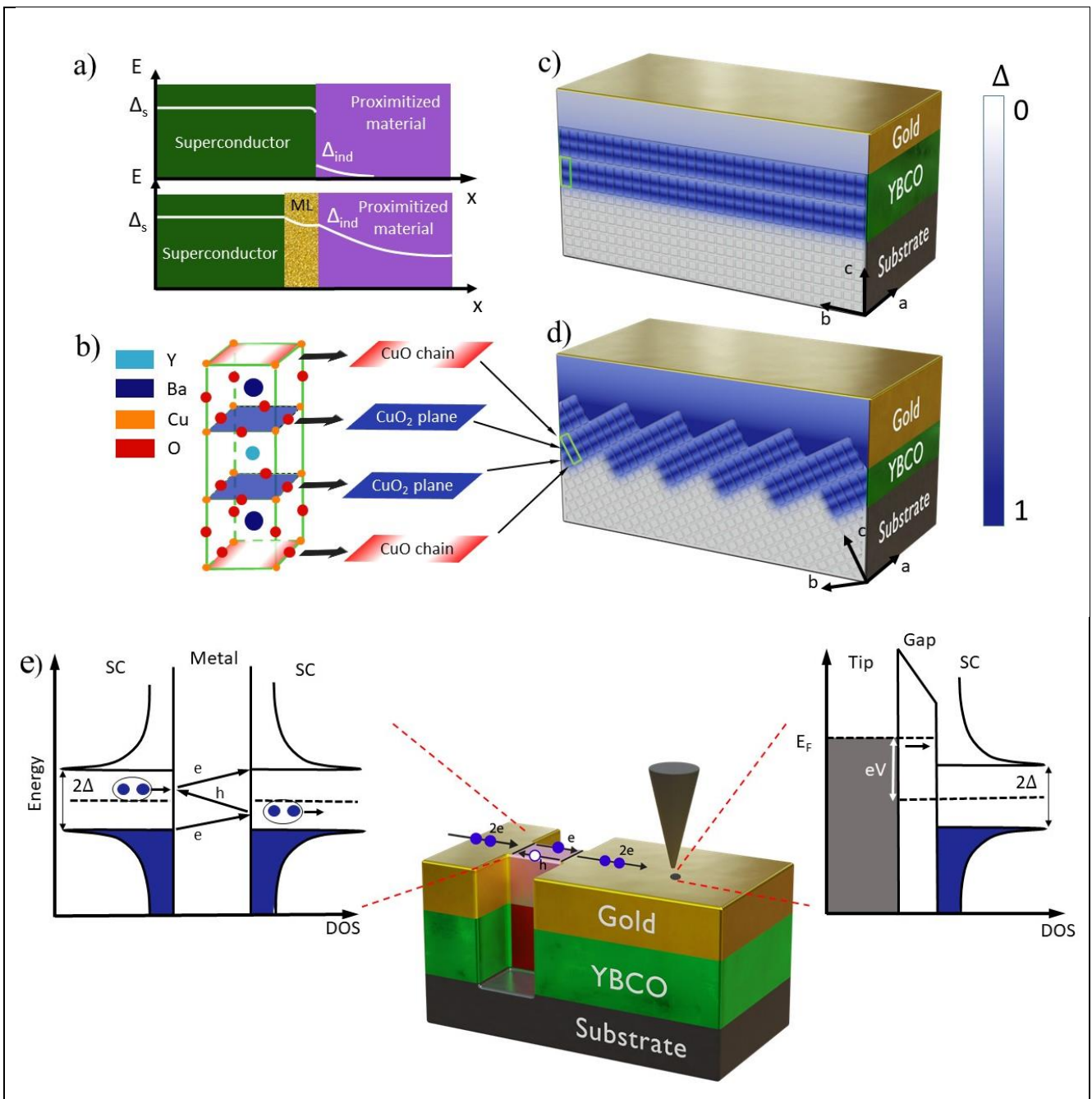

**Figure 1**. Au/YBCO electrodes. (a) Evolution of the order parameter in the hybrid heterostructure, where ML is a mediating layer, $\Delta_s$ and $\Delta_{ind}$ are an energy gap in a superconductor and an induced energy gap, respectively. (b) YBCO unit cell. (c) Sketch of the conventional hybrid high-$T_c$ electrode, where the c-axis is normal and the $CuO_2$ planes are parallel to the film surface. (d) Sketch of the hybrid device based on vicinal substrates, where the c-axis is inclined towards the substrate surface normal. A schematic illustration of the order parameter variation in (c) and (d) is shown in a scale bar on the right side. (e) A schematic representation of an STM (right side) and MAR spectroscopy (left side) used to study induced order parameter in gold. Normal-state areas in the constriction area of YBCO and gold layers are indicated in red and pink, respectively. The black arrow in the right panel shows a tunneling current. SC is a superconductor.

**Concept.** Here, we propose an alternative approach to the fabrication of the high-$T_c$ superconducting electrodes for hybrid devices based on superconducting films deposited on vicinal substrates. The **c**-axis of the film grown on the vicinal substrate is inclined towards the substrate surface normal (Figure 1d). The superconducting $CuO_2$ planes have direct access to a film surface providing better conditions for the injection of Cooper pairs into the material where induced superconductivity is required. Such a configuration is expected to have a larger energy gap and smaller spatial variation of the order parameter at the superconducting film surface. In our work, we use gold as the mediating layer because it meets the abovementioned requirements and has a large normal coherence length $\xi_n$ resulting in a reduced spatial variation of the order parameter.[27] The sketch of the novel hybrid superconducting Au/YBCO electrode with the gold mediating layer is shown in Figure 1d.

We perform a comprehensive study of this hybrid superconducting Au/YBCO electrode using STM and multiple Andreev reflection (MAR) spectroscopy, as shown in Figure 1e. The STM provides information on the density of states at the surface of the studied material with nanometer spatial resolution (right panel in Figure 1e), while the MARs in the nanoconstriction conductance (left panel in Figure 1e) are an indicator of the superconducting nature of the observed energy gap.

**Results and discussion**

**STM measurements.** We fabricated a number of Au/YBCO heterostructures consisting of 30-35-nm-thick YBCO films deposited on vicinal $SrTiO_3$ (STO) and $NdGaO_3$ (NGO) substrates and covered *in situ* by a 15-nm-thick gold layer. Figure 2a shows a representative surface STM topography of the Au/YBCO heterostructure fabricated on a (110) NGO substrate with 10.5° miscut. Along with a granular structure of the gold film with an average grain size of 10 nm, elongated parallel terraces in the direction of the red arrow are observed. These terraces are formed due to the step-flow growth of the YBCO films on the vicinal substrates. To avoid a large step-bunching[28] during the step-flow growth and achieve a low roughness of the Au/YBCO electrode surface within a few unit cells, we optimized the substrate surface treatment and sputtering parameters.[29, 30] The corresponding root-mean-square roughness of the Au/YBCO electrode surface along and perpendicular to the terraces is 0.29 and 0.84 nm, respectively. A scanning electron micrograph (SEM) of an Au/YBCO electrode cross-section is shown in the inset of Figure 2a. The analysis of the cross-section confirms the thickness of the gold layer and shows that its surface topology follows that of the YBCO film.

We performed differential conductance spectroscopy (d$I$/d$V$) at a temperature $T = 10$ K following a line between two terraces (see Methods for more details on the STM measurements) in a span of 20 nm (inset in Figure 2b). To avoid oxygen depletion in the YBCO, we did not anneal the sample before measurements. The d$I$/d$V$ spectra show an energy gap varying from 10 to 17 meV at the surface of the gold layer compatible with proximity-induced superconductivity (more examples see in Supplementary Figure S1). Neither a zero-bias conductance peak due to the nodal quasiparticles nor distinct coherence peaks were observed. Moreover, spectra show no significant correlation of the induced energy gap value with the gold grain size, expected for a situation when the Coulomb blockade is the origin of the observed energy gap.[31] We determine the value of the energy gap as half of the distance between a dip and kink values of the spectra in Figure 2b following the work of Stepniak *et al*,[32] where STM measurements using tungsten and niobium tips are compared. The energy gap value may vary at the gold film surface because the normal coherence length $\xi_n$ in gold is comparable with the terrace width at the measurement temperature T=10 K.

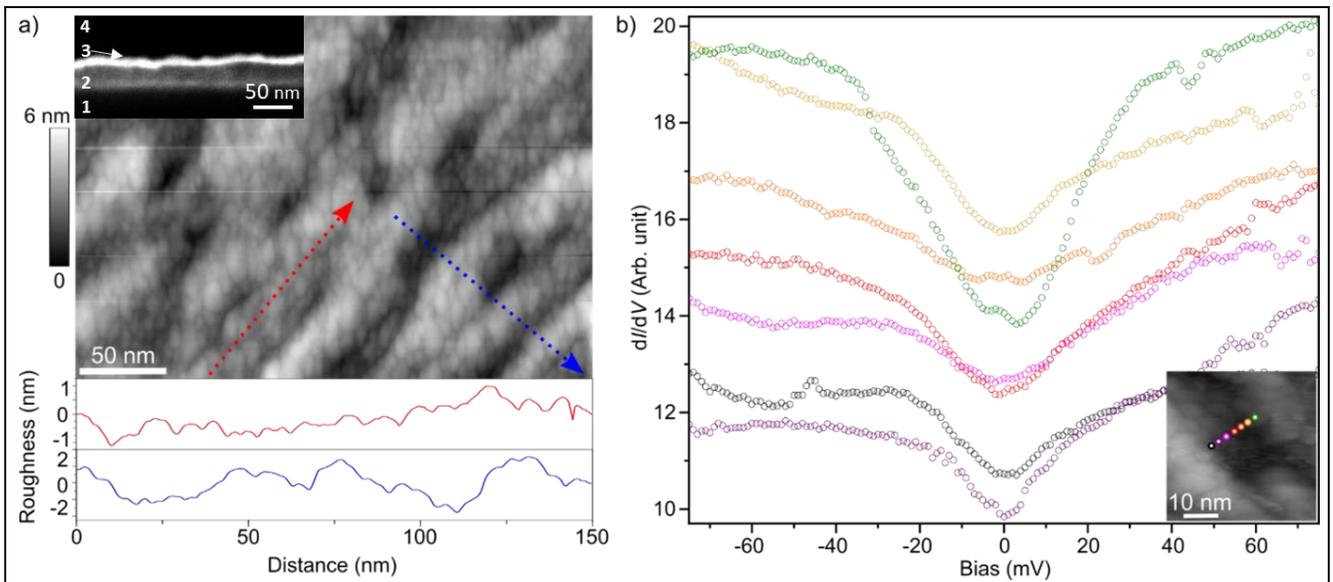

**Figure 2.** Surface morphology and spectroscopic properties of the Au/YBCO surface. (a) Scanning tunneling micrograph of an Au/YBCO heterostructure on a vicinal NdGaO substrate (upper panel) and surface profiles along and perpendicular to the terraces (red and blue dotted arrows) ($V_{set}$ = 1V, $I_{set}$ = 130 pA). Inset shows a scanning electron micrograph of an Au/YBCO heterostructure cross-section. The substrate, YBCO film, Au film, and electron beam deposited Pt film are numbered 1, 2, 3, and 4, respectively. (b) Differential conductance spectra taken at different positions of the sample. The spectra are offset vertically for clarity. Inset shows the measurement points ($V_{set}$ = 100 mV, $I_{set}$ = 2 nA, $V_{mod}$ = 1 mV). The colour of each dot corresponds to the colour of differential conductance spectra taken at this point.

The anisotropic pairing symmetry in high-$T_c$ cuprate superconductors makes the interpretation of the density of states more complicated in comparison to BCS superconductors.[33] Sometimes it leads to the question of whether the observed energy gap in the density of states is due to superconductivity. For example, recent ARPES and STM studies of the topological insulator $Bi_2Se_3$ grown on BSCCO show discrepancies[16-21] which have been recently resolved. The apparent energy gap is the consequence of the dynamic Coulomb blockade.[34] However, in our work we exclude Coulomb blockade as the origin of the observed gap since our measurements are performed in a continuous 15-nm-thick gold film, as opposed to metallic nanoislands or disordered atomically thin wetting layers.[31, 35, 36] To study the origin of the energy gap and get further insight into the physics of the induced superconductivity in the Au/YBCO electrodes, we fabricated Au/YBCO nanoconstrictions and investigate their electrical properties.

**MAR spectroscopy.** The Andreev reflection is a phenomenon specific to the charge transfer through the normal metal-superconductor interface. A normal electron with energy below the superconducting gap energy $\Delta_S$ is retroreflected as a hole from this interface creating a Copper pair at the Fermi level of the superconductor and providing an additional charge transfer compared to the electrons with energies above $\Delta_S$. Within the framework of the Blonder-Tinkham-Klapwijk model,[37] a nanoconstriction can be considered as a superconductor-normal metal-superconductor (SNS) junction where the voltage at currents above the critical current is developed across the dissipative neck region. According to the theoretical approach to the SNS junction, each quasiparticle undergoes MAR before it is scattered or leaves the pair potential well, as shown schematically in Figure 1e. If a quasiparticle undergoes n Andreev reflections, then *ne* charges are transferred through a normal metal-superconductor boundary in addition to the initial one, and the SNS junction current is enhanced due to Andreev reflection. Here, n is an integer and e is an electron charge. As a consequence of MAR, conductance curves show a series of features at voltages $V = 2\Delta/ne$. The measurements of the Andreev reflection spectrum give clear evidence of the superconducting nature of the energy gap and provide information on its magnitude.[38, 39]

60-495-nm-wide and 40-60-nm long Au/YBCO nanoconstrictions were fabricated along and perpendicular to the terraces formed by the step-flow growth with focused ion-beam milling or electron-beam lithography with ion-beam etching (see Methods). The constrictions were made as short as possible to reduce the scattering in the constriction neck area. Nanoconstrictions patterned along the terraces had high critical current density $J_c=I_c/Wd$ up to 82 MA/cm$^2$, confirming that the YBCO does not degrade after the nanoconstriction patterning. Here, *W* is the width of a nanoconstriction, *d* is the thickness of a YBCO film. In this work, we study the electrical transport of the nanoconstrictions patterned across the terraces, which have several times lower critical current densities to avoid overheating at high voltage biases. Figure 3a and the inset in Figure 3a show a representative SEM image of a microbridge with a nanoconstriction and its zoomed image, respectively.

A representative resistance temperature dependence *R*(*T*) (the inset in Figure 3b) has a two-step transition as a result of the weak-link behaviour of a nanoconstriction. The first resistance drop with the onset at 90.5 K occurs due to the superconducting transition of the large-size electrodes of the nanoconstrictions. The constriction itself switches to the zero-resistance state at the lower temperature around $T = 81$ K when the thermally-activated phase slippage is suppressed.

A representative current-voltage (*IV*) curve of a 100-nm-wide nanoconstriction is shown in Figure 3b. The nanoconstriction has a normal-state resistance $R_n = 14.5$ Ω and a critical current $I_c$=176 μA at a temperature of 4.2 K. Having linear dependence above the critical current, the *IV* curve demonstrates neither voltage steps characteristic of phase slippage,[40-42] nor power law dependence $V \sim I^\alpha$ due to flux flow.[40] Thus, we conclude that the appearance of the normal-state region in the nanoconstriction neck is a dominant mechanism of the resistive state at currents above the critical current. The size of this region is stabilized due to the high thermal conductivity of the gold film. Therefore, our constrictions can be considered as an SNS junction at a current above the critical current. A higher value of an excess current in comparison to the critical current, as shown in Figure 3b by the dashed line, confirms the high transparency of the interfaces normal-state region/superconductor (red in the inset of Figure 4a) and normal-state region/gold (pink in the inset of Figure 4a) due to the matching of Fermi velocities.[43]

For MAR spectroscopy we chose bridges with the width ranging from 60 to 130 nm. A representative conductance curve of the 100-nm-wide nanoconstriction measured at $T = 4.2$ K shows a nonmonotonic step-like behavior with a number of conductance steps with corresponding dips (Figure 4a). Here, we follow the numerical simulations of

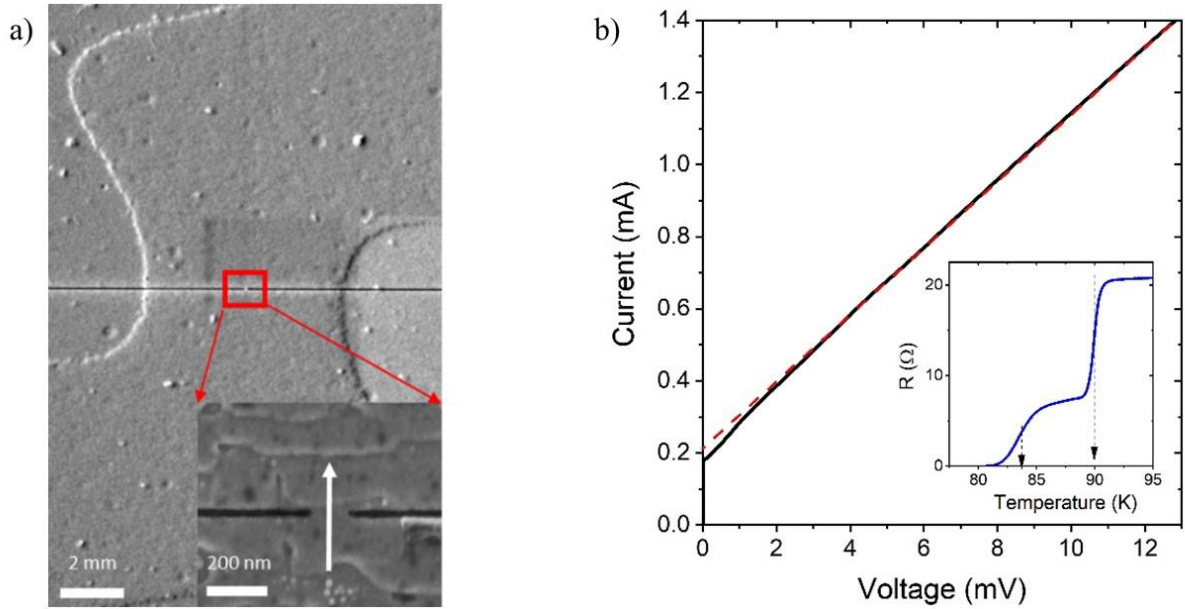

**Figure 3**. Structural and electrical characteristics of the constriction. (a)Space SEM image of a microbridge with a nanoconstriction made by FIB. Inset shows the nanoconstriction with a width of 100 nm, the current flows perpendicular to the terraces and is indicated by the white arrow (b) Typical *IV*-curve of a nanoconstriction with a width of 100 nm, the extrapolation by dashed red line shows an excess current. The inset shows a typical curve of a transition from normal to superconducting state; the transition temperatures of the electrode and nanoconstriction are shown by dashed arrows.

Popovich *et al.*[44] for the SNS junctions with transparent normal metal-superconductor interfaces to assign the position of the conductance dips with the $n^{th}$ MAR harmonic at $V = 2\Delta/ne$. The position of a smeared dip at $V = 41.2$ mV indicated by the green arrow is close to the expected values of $\Delta/e$ corresponding to the energy gap in the **b**-axis direction of YBCO $\Delta_b = 44$ meV.[45] Although this nanoconstriction demonstrates one conductance dip due to the intrinsic energy gap in YBCO, one should take into account that YBCO films are twinned and one can observe the conductance dips that belong to the energy gaps either in one antinodal direction $\Delta_a$ or $\Delta_b$ or both of them (see Supplementary Figure S2). The temperature dependence of $\Delta_b$ energy gap is close to the Bardeen-Cooper-Schrieffer (BCS) theory prediction, which is shown in Figure 4b by the black dashed line. In addition, Figure 4a shows a series of the low-voltage conductance dips indicated by the dark-yellow arrows. Their positions are in perfect agreement with the $V_n = 2\Delta_1/ne$ dependence with $\Delta_1 = 14$ meV shown by the black line in the insert in Figure 4b for integer n from 2 to 5. The temperature dependence of this gap is different from that of the intrinsic YBCO energy gap (Figure 4b). We assign the energy gap with $\Delta_1(4.2K) = 14$ meV to the induced superconducting gap in the gold layer. The magnitude of this gap ranging from 11.25 to 15.3 meV for different nanoconstrictions is close to that measured with STM at the gold layer surface. The ratio between the induced and corresponding intrinsic superconducting energy gap is ranging from 0.34 to 0.41 (see Supplementary Information).

Moreover, we observe a pronounced dip at $V = 5$ mV with temperature dependence different from the abovementioned dips with energy gaps of $\Delta_b(4.2K) = 41.2$ meV and $\Delta_1(4.2K) = 14$ meV. This dip at voltages ranging from 2.5 to 5 meV is systematically observed for many samples. To exclude the second weak link, which may appear in the constriction area, as an origin of this dip, we compare the temperature dependences of the critical current $I_c$ and this dip and find that they are different. Therefore, we assign this conductance dip to the $\Delta_2$ energy gap. On the one hand, it was theoretically predicted that so-called surface states appear at the atom layer close to the surface of a metal film.[46] Later, Wei et al.[47] experimentally showed that a superconductor can induce bulk and surface energy gaps in gold. Then the dip with $\Delta_2(4.2K) = 5$ meV may correspond to the induced energy gap in the surface states of gold. On the other hand, the amplitudes of the conductance steps at $V = 5$ mV and $V = 14$ mV are close. Therefore, the $\Delta_2$ energy gap may belong to the bulk gold and can be induced by the YBCO energy gap in **c**-axis direction.[48, 49] To clarify the nature of this gap, further investigations are required.

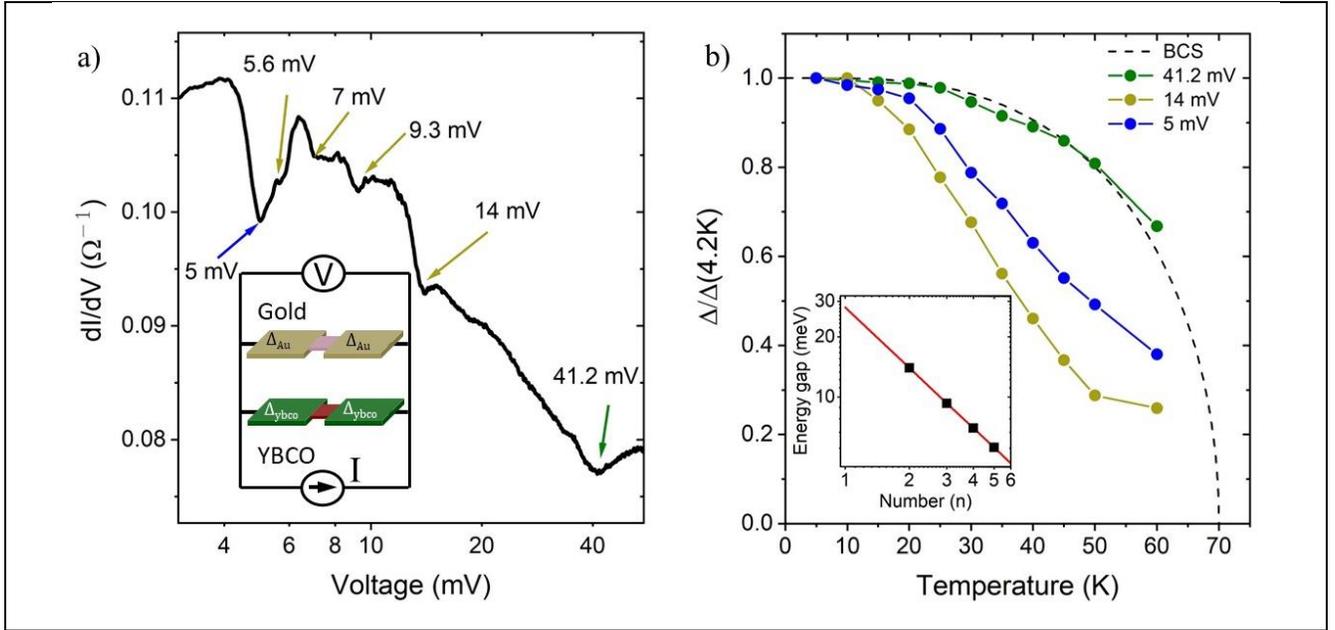

**Figure 4**. Andreev spectroscopy of the superconducting energy gaps in Au/YBCO heterostructure. (a) Conductance of a 100-nm-wide Au/YBCO nanoconstriction at $T = 4.2$ K. In the inset, an Au/YBCO constriction is presented as a parallel connection of two SNS junctions. A normal state area of the constrictions at the current above the critical current for YBCO and gold is shown in red and pink, respectively. (b) Temperature dependence of the normalized energy gaps in Au/YBCO heterostructure. The BCS model prediction is shown by the dashed black line. Inset shows the positions of the conductance dips indicated by dark-yellow arrows. The solid red line represents $V_n = 2\Delta_1/ne$ dependence with $\Delta_1 = 14$ meV.

Having obtained the value of the induced energy gap with two independent techniques, we would like to estimate the induced superconducting gap theoretically. Sharoni et al.[26] studied the proximity effect in gold films atop the **c**-axis oriented YBCO films and found that the induced superconducting energy gap exponentially decays over the thickness of the gold film as $\Delta_{Au}(d) = \Delta_0 e^{-d/\xi_n}$, where $\Delta_0 = 15$ meV is the superconducting energy gap at the Au/YBCO interface, $\xi_n$ is the normal coherence length in metal, and $d$ is the distance from the **a**-axis YBCO surface. If such an exponential decay was in our system, we would observe smeared MAR conductance steps. However, it contradicts our experimental data.

In order to obtain a rough check of our results as compared to other theoretical considerations, we resort to the McMillan model of tunneling between a superconductor and a normal metal where the homogeneous induced order parameter in normal metal is assumed.[50] Within the framework of this model, the energy gap $\Delta_n$ induced in the normal metal film with thickness $D_n$ by the superconducting film with thickness $D_s$ and energy gap $\Delta_s$ can be estimated as $\Delta_n = \Delta_s/[1+\Gamma_s/\Gamma_n]$ (1), where $\Gamma_n$ and $\Gamma_s$ are scattering rates in the normal and superconducting films, respectively. First of all, the McMillan's model assumes a tunneling barrier at the normal metal-superconductor interface. The large mismatch in Fermi velocities of gold $v_F(Au) = 1.4 \cdot 10^6$ m/s[51] and YBCO $v_F(YBCO) = 2 \cdot 10^5$ m/s[52] favors the applicability of the McMillan's model in terms of tunneling. Within the simplified ballistic approach, the scattering rate in the gold film can be roughly estimated as $\Gamma_{Au} \approx \hbar v_F(Au)/2D_{Au}$, where $\hbar$ is the reduced Plank's constant and $D_{Au} = 15$ nm is the gold film thickness. From available experimental data for cuprate superconductors BSCCO and LSCO,[53, 54] we find the scattering rates in the surface layer of these superconductors to be roughly equal $\hbar v_F/\xi$, where $\xi$ is the in-plane coherence length. Then, due to the large scattering rate, once the quasiparticle crossed the Au/YBCO interface from Au into YBCO, it has a high probability of being reflected back to Au while scattering in the superconductor layer with a thickness of the order of $\xi$. Therefore, we roughly estimate the scattering rate in YBCO as $\Gamma_{YBCO} \approx \hbar v_F(YBCO)/2\xi_{ab}$, where $\xi_{ab} = 1.3$ nm is the in-plane coherence length in YBCO.[55] Substituting the abovementioned values of $\Gamma_{Au}$ and $\Gamma_{YBCO}$ in equation (1) we obtain $\Delta_{Au} \approx \Delta_{YBCO}/[1+(D_{Au}v_F(YBCO)/\xi_{ab} v_F(Au))] \approx 11 - 17$ meV (2) which is in good agreement with our experimental results. Here, $\Delta_{YBCO} = 29 – 44$ meV is the superconducting energy gap in YBCO.[45] While the assumptions for the McMillan model are not all perfectly fulfilled in our case, it shall nevertheless serve as an estimate of the energy scales we observe in experiment. To further demonstrate the consistency of our model, we fit the values of the induced energy gap measured by Sharoni at al.[26] at the surface of the gold film above the **a**-plane facets of YBCO with Eq. 2. Here, we use $\Delta_{YBCO}$ and $\xi_{ab}$ as free parameters and obtain $\Delta_{YBCO} = 25\pm4$ meV and $\xi_{ab} = 1.1\pm0.3$ nm which are close to the values of the energy gap in the **a**-axis direction and the in-plane coherence length of YBCO, respectively.

Due to the large scattering rate in YBCO close to the Au/YBCO interface, the layers with the intrinsic and induced energy gaps are spatially separated. Therefore, we can model Au/YBCO constrictions as a parallel connection of two SNS junctions where the voltage at currents above the critical current is developed across the dissipative neck region connecting both YBCO and gold electrodes as shown in the inset in Figure 4a. It allows us, in contrast to conventional SNS junctions, to observe both intrinsic and induced energy scales. The higher number of Andreev reflections observed for the gold part of the nanoconstrictions can be explained by weaker scattering in gold compared to YBCO. Therefore, we propose a novel technique to study induced superconductivity using Andreev reflection spectroscopy of multi-layer nanoconstrictions. Such a technique could be extremely helpful for objects such as hybrid structures capped by an insulating layer or topological insulators with a bottom superconducting electrode that are difficult to access with conventional surface-sensitive techniques.

The fabrication of hybrid devices sometimes requires elevated temperatures which can lead to the YBCO degradation. For example, the fabrication of a YBCO/Au/graphene hybrid device using the standard stamp technique includes the heating of the device up to 180°C to remove the stamp. To prove the ability of the Au/YBCO electrode to withstand this temperature, we anneal the Au/YBCO electrodes for 10 min at a temperature of 180°C in oxygen at a pressure of 0.8 bar and find that the change in the critical temperature of the electrodes is 1 K only.

In summary, we have fabricated and investigated the Au/YBCO heterostructures on vicinal substrates. STM measurements show an energy gap ranging from 10 to 17 meV at the surface of the 15-nm-thick gold layer. To study the origin of the energy gap in the gold layer, we fabricate the nanoconstrictions from the Au/YBCO heterostructure and perform Andreev reflection spectroscopy which confirmed the STM results. To the best of our knowledge, the obtained values of the induced energy gap at the surface of the hybrid Au/YBCO heterostructures are the largest ever observed confirming the promise of the new approach to the fabrication of hybrid devices with high-$T_c$ superconducting electrodes. Low roughness, large induced energy gap, and good compatibility with different proximitized materials make the Au/YBCO electrodes a versatile platform for the development and investigation of hybrid devices in a broad temperature range. We consider Andreev reflection spectroscopy with hybrid nanoconstrictions to be a promising technique to determine the induced energy gap in emerging 1D and 2D materials including topological insulators.

## Methods

**Heterostructure fabrication.** The epitaxial YBCO films were deposited by dc sputtering at a high oxygen pressure of 3.4 mbar on vicinal SrTiO$_3$ or NdGaO$_3$ substrates with a **c**-axis inclined by 8 or 10.5 degrees towards the substrate surface normal. Before the sputtering, the substrates were etched with a buffered oxide etch for TiO$_2$ or GaO$_2$ surface termination, respectively. The temperature of the heater was 950 °C during the YBCO film deposition. A deposition rate of 1 nm/minute was calculated from profilometer measurements. After the deposition of a 30-35 nm film, the heater temperature was lowered to 550 °C and the film was annealed for 25 min in pure oxygen at the pressure of 800 mbar. Then the heater temperature was ramped down to room temperature. In the next step, the substrate with the YBCO film was transferred into another sputtering chamber and a 15 nm-thick gold film was deposited *in situ* by dc magnetron sputtering of the gold target in argon at the pressure of $5 \cdot 10^{-3}$ mbar. The temperature of the heater was 90 °C during the gold film deposition. 100-nm-thick gold contact pads were deposited *ex situ* by dc magnetron sputtering at room temperature.

**Nanoconstriction fabrication.** The nanoconstrictions were fabricated with a two-step process. In the first step, the Au/YBCO structure was patterned into 13 microbridges with a width of 6 μm by the UV- lithography technique followed by ion beam etching in argon and then etching in Br-ethanol solution. The microbridges were oriented along and perpendicular to the terraces formed by the step-flow growth. 40-60-nm-long nanoconstrictions with widths ranging from 100 to 495 nm were fabricated using focused ion beam milling. To protect a sample from Ga ions while milling, it was covered by a 30-nm-thick layer of PMMA resist, and then a 90 nm-thick layer of gold deposited by dc magnetron sputtering. The protection layer was removed in acetone after the nanopatterning. The details of FIB milling are available elsewhere.[42] The nanoconstrictions with a width below 100 nm were made using inverse process. In the first step, the nanostructures were fabricated by the electron-beam lithography using the ion-beam etching in argon through the 80-nm-thick CSAR62 resist mask. In the second step, the microbridges were defined by the UV-lithography and wet chemical etching using the alignment markers fabricated during the first step.

**Experimental setup.** The electrical characteristics of the current-biased nanoconstrictions were measured by a four-probe technique inside a Dewar insert in a 4-80 K temperature range. The differential resistance of the nanoconstriction was measured with a lock-in amplifier at a modulation frequency of 10 kHz. The temperature of the sample was maintained by a resistive heater controlled by a Lakeshore 335 temperature controller.

**STM measurements**. Scanning tunneling data were acquired in a commercial Createc system in an ultra-high vacuum at a base pressure of $P < 10^{-10}$ mbar and a base temperature of 10 K using a tungsten tip. Tunneling spectra were acquired using standard lock-in techniques at $f = 767$ Hz and $V_{mod} = 1$mV. The tungsten tip was chemically etched and flashed at high temperatures to remove any impurities. The Au/YBCO sample was transported in ambient conditions and introduced in UHV after pumping the load lock chamber overnight. No annealing in UHV was performed to avoid thermal degradation of YBCO.

**Acknowledgments**

We thank S. Droege for his help in preparing the illustrations, Y. Starmans for his assistance in the measurements, D. Rohe for fruitful discussions, S. Trellenkamp, and F. Lentz for their help with electron-beam lithography. The FIB nanopatterning was performed at Ernst-Ruska Center of Forschungszentrum Jülich within the project FZJ-PGI-9-LM1. I.G. was supported by the Deutsche Forschungsgemeinschaft (DFG, German Research Foundation) under Germany´s Excellence Strategy – Cluster of Excellence Matter and Light for Quantum computing (ML4Q) EXC 2004/1 390534769. J.M.C. acknowledges funding from the Alexander von Humboldt Foundation.

**Additional information**

The authors declare no competing interests.

**Supplementary Information**

**STM measurements**

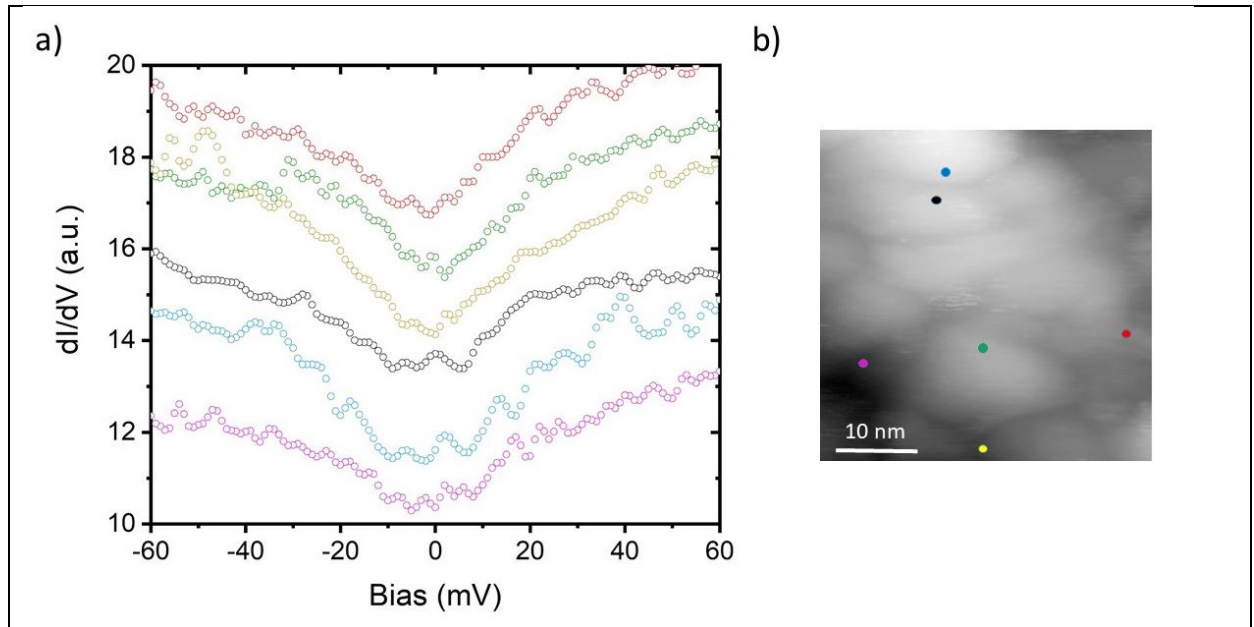

**Figure S1.** STM measurements. (a) Differential conductance spectra taken at different positions of the sample. The spectra are offset vertically for clarity. (b) The map of the measurement points ($V_{set}$ = 80 mV, $I_{set}$ = 2 nA, $V_{mod}$ = 1 mV). The colour of each dot corresponds to the colour of differential conductance spectra taken at this point.

**Multiple Andreev reflections in Au/YBCO nanoconstrictions**

YBa$_2$Cu$_3$O$_{7-x}$ (YBCO) possesses d-wave symmetry of the order parameter with different superconducting energy gaps in the **a**-axis and **b**-axis directions, $\Delta_a$ =29 meV and $\Delta_b$ =44 meV.[45] One can observe conductance steps corresponding either to the superconducting energy gap $\Delta_a$ or $\Delta_b$ due to the twinning of the YBCO films. The simultaneous appearance of the conductance steps related to both superconducting energy gaps $\Delta_a$ and $\Delta_b$ in Figure 2a may be explained by the broader angle of trajectories for wider nanoconstrictions. We assume that the energy gaps in gold $\Delta_a^{Au}$ = 11.5 meV and $\Delta_b^{Au}$ = 15.3 meV identified in Figure 2a are induced by corresponding energy gaps in YBCO $\Delta_a$ = 28 meV and $\Delta_b$ = 42 meV.

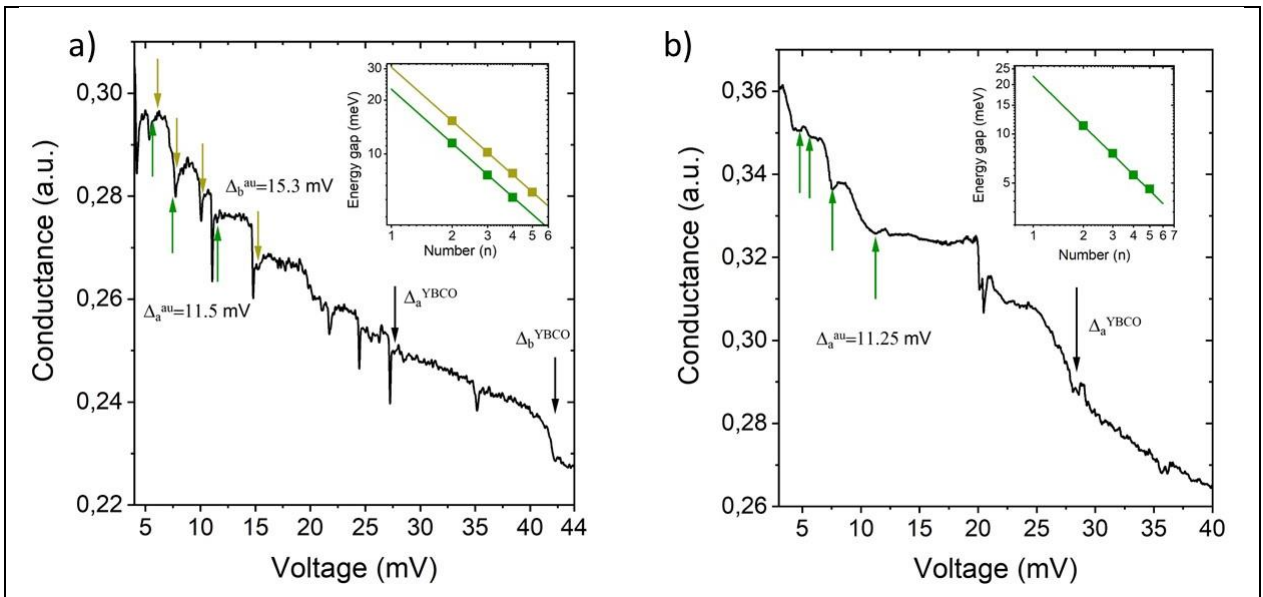

**Figure S2.** Andreev spectroscopy of the superconducting energy gaps in Au/YBCO heterostructure. (a) Conductance of a 90-nm-wide Au/YBCO nanoconstriction at the temperature of $T = 4.2$ K. Inset shows the positions of the conductance dips indicated by green and dark-yellow arrows, respectively. The solid green and dark-yellow lines represent $V_n = 2\Delta/ne$ dependence with $\Delta_a^{Au} = 11.5$ meV and $\Delta_b^{Au} = 15.3$ meV, respectively. (b) Conductance of a 60-nm-wide Au/YBCO nanoconstriction at $T = 4.2$ K. Inset shows the positions of the conductance dips indicated by the green arrows. The solid green line represents $V_n = 2\Delta/ne$ dependence with $\Delta_a^{Au} = 11.25$ meV.